# What can we learn by squeezing a liquid

R.Casalini[1,2,@], S.Capaccioli[3,§] and C.M.Roland[1,*]

[1] *Naval Research Laboratory, Code 6120, Washington DC 20375-5342*
[2] *George Mason University, Fairfax VA 22030*
[3] *Dipartimento di Fisica and INFM, Università di Pisa, Pisa, Italy* and *CNR-INFM, CRS SOFT, Università di Roma "La Sapienza", Piazzale Aldo Moro, Roma, Italy.*

[@] e-mail: casalini@nrl.navy.mil, [*] e-mail: roland@nrl.navy.mil,
[§] e-mail: capacci@df.unipi.it

**Abstract**

Relaxation times $\tau(T,\upsilon)$ for different temperatures, $T$, and specific volumes, $\upsilon$, collapse to a master curve versus $T\upsilon^\gamma$, with $\gamma$ a material constant. The isochoric fragility, $m_V$, is also a material constant, inversely correlated with $\gamma$. From these we obtain a 3-parameter function, which fits accurately $\tau(T,\upsilon)$ data of several glass-formers over the supercooled regime, without any divergence of $\tau$ below $T_g$. Although the 3 parameters depend on the material, only $\gamma$ significant varies; thus, by normalizing material-specific quantities related to $\gamma$, a universal power law for the dynamics is obtained.

By cooling a liquid in a time shorter than the crystallization time, a metastable equilibrium is reached, referred as a supercooled state. As effectively, and in fact more rapidly, a liquid can be supercooled by squeezing (i.e. by applying hydrostatic pressure). Herein we discuss how the fundamental difference between these two thermodynamical routes to vitrification can yield insights into the physics behind this metastable state.

Certainly the most intriguing phenomenon observed on cooling supercooled liquids is the dramatic slowing down of the structural dynamics, by more than 14 orders of magnitude over a relative small range of temperature. Eventually the structural relaxation time $\tau$ becomes so large that the molecular motion ceases, at least on the experimental time scale. Thus, macroscopically the system behaves as a solid, even though no apparent changes have transpired in its microscopic structure. Gaining a molecular-level understanding of this phenomenon is considered to be one of the most challenging problems of condensed matter physics.

In a typical experiment a sample is cooled at some rate and below a certain temperature (the glass temperature $T_g$) this cooling rate becomes comparable with $\tau^{-1}$, whereby the system cannot attain thermodynamical equilibrium in the time of measurement. Consequently, $T_g$ is rate dependent and its definition is somewhat arbitrary. Typically for dielectric relaxation measurements $T_g$ is taken such that $\tau(T_g)=100$s (as used herein), while for viscosity measurements $\eta=10^{12}$Pa s at $T_g$.

A classification of the effect of temperature on the dynamics is the fragility or "steepness index",[1-3] defined as $m_P = \partial \log(x) / \partial (T_g/T)\big|_{T=T_g, P=const}$, where $x$ can be $\tau$ or $\eta$. This parameter varies for small molecules and polymers in the range $35 \leq m \leq 214$,[4-7] which reveals the drastic differences in the dynamic slowing down of different materials. However, currently there is no accepted molecular-level interpretation of the origin of fragility.

The two pathways to $T_g$ (cooling and compression) have an interesting difference: an isobaric change of $T$ alters both $V$ and $T$ while an isothermal change of $P$ only affects $V$. Consequently, using high pressure measurements it is possible to deconvolute the relative effect of $T$ and $V$ on the dynamics, which is of fundamental importance in assessing theoretical models and their foundation.

High pressure experiments over the past few years on an extended number of samples[8-11] have established unambiguously that both temperature and volume govern the temperature dependence of the relaxation time, at least at atmospheric pressure. Therefore a complete model of the glass transition should be able to provide an equation for the dependence of the relaxation time on both temperature and volume. Recently, it has been shown[12-15] that the behavior of $\tau(T,\upsilon)$ can be rescaled onto a master curve when plotted versus $T\upsilon^\gamma$.

$$\tau(T,\upsilon) = \Im(T\upsilon^\gamma) \qquad (1)$$

where $\upsilon$ is the specific volume and $\gamma$ a material constant. Therefore the relevant variable to describe structural relaxation times is the product $T\upsilon^\gamma$.

This scaling was first observed for ortho-terphenyl (OTP)[16,17] for $\gamma = 4$ (the value predicted for a simple Lennard Jones 6-12 potential[18,19]), and then shown to be generally valid for many others materials with $0.13 \leq \gamma \leq 8.5$.[12-15] Equation (1) can describe dielectric,[12-15] light scattering[17,20] and viscosity measurements,[21] with comparable values for the constant $\gamma$ obtained. The scaling behavior has also been confirmed in simulations.[22]

The validity of the scaling (eq.(1)) provides a straightforward deconvolution of the effects of $T$ and $\upsilon$. If the behavior of the relaxation time at constant pressure is known, it is possible to determine the behavior at constant volume from the equation of state $\upsilon(T,P)$ and the value $\gamma$ for that particular glass-former.[23] Defining the isochoric fragility as $m_V = \partial \log(x)/\partial(T_g/T)\big|_{T=T_g, V=const}$, from eq.(1) it follows that $m_P$ can be calculated from $m_V$ as [12,13]

$$m_P = m_V\left(1 + \gamma\alpha_P T_g\right) \qquad (2)$$

where $\alpha_P$ ($\equiv 1/V\, \partial V/\partial T\big|_P$) is the isobaric (volume) expansion coefficient at $T_g$. Actually, this relation is valid for any $T$ as long as $\alpha_P$ is determined at the same $T$ ($\gamma$ is constant). Very generally, $0 \leq m_V \leq m_P$, with $m_V$ being smaller than $m_P$ since the former does not include the effect of volume changes. The two limiting cases $m_V = m_P$ and $m_V = 0$ correspond to $\tau$ being an unique function of $T$ and $V$, respectively.

In addition to eq.(2), there are two other independent relations: Comparing the value of $m_V$ and $m_P$ for more than 38 glass formers: (i) the two fragilities are found to be linearly correlated, $m_P = (37 \pm 3) + (0.83 \pm 0.05) m_V$; and at the same time (ii) $m_V$ is also related to $\gamma$ according to $\gamma = -1.042 + 217/m_V$.[24]

The consequence of these two correlations is the unintuitive result that the temperature behavior, for both the isobaric and the isochoric conditions, is related to the parameter $\gamma$ which (viz. eq.(1)) describes the dependence of the relaxation time on volume. Thus, the scaling exponent (whose physical interpretation has been discussed previously[12,25]) serves as a "new" metric to classify glass formers in terms of their fragility. Indeed, herein we go a step further and identify from $\gamma$ a variable which unifies the behavior of all glass-formers.

An equation has been found which best describes (with the fewest parameters) the $\tau(T,\upsilon)$ behavior over temperatures and volumes ranging from the glass transition up to temperature below that at which the behavior becomes activated[26]

$$\log[\tau(T,V)] = \log(\tau_\infty) + B\left(\frac{T_g \upsilon_g^\gamma}{T\upsilon^\gamma}\right)^D \qquad (3)$$

where $\tau_\infty$, $B$ and $D$ are constants with $\tau_\infty$ the limit value of $\tau$ at high temperatures. This equation can be derived from a model relating the dynamics to the system entropy.[26] Assuming a linear relationship between $m_V$ and $m_P$, $m_P = a + bm_V$ with $a$ and $b$ constants. Then for $m_V = 0$, $m_P = m_P^{min} = a$, while $m_V = m_P = m_P^{max}$, and hence it follows that $b = 1 - m_P^{min}/m_P^{max}$. We can then rewrite the linear relationship between $m_P$ and $m_V$ as

$$m_P = m_P^{min} + \left(1 - \frac{m_P^{min}}{m_P^{max}}\right) m_V \qquad (4)$$

Combining eq.(2) with eq.(4) it follows that

$$\gamma = \frac{m_P^{min}}{\alpha T_g}\left(\frac{1}{m_V} - \frac{1}{m_P^{max}}\right) \qquad (5)$$

Comparing eq.(5) with the phenomenological correlation found between $\gamma$ and $m_V$ we obtain $\left(m_P^{min}/m_P^{max}\right) = 1.04 \times \alpha T_g$. Thus, to a good approximation

$$\frac{m_P^{\min}}{m_P^{\max}} \cong \alpha T_g \qquad (6)$$

Eq.(6) agrees with results from the phenomenological correlation between $m_V$ and $m_P$, $m_P^{\min} = 37 \pm 3$ and $m_P^{\max} = 231 \pm 72$.[24] From eq.(6) $\alpha T_g = \frac{m_P^{\min}}{m_P^{\max}} = 0.16 \pm 0.06$, in good agreement with the literature; e.g., the Boyer-Bondi rule $\alpha T_g = 0.16 \div 0.19$.[27] Accordingly, we substitute eq.(6) in eq.(5) finding

$$\gamma = \frac{m_P^{\max}}{m_V} - 1 \qquad (7)$$

Defining $m_P^{\infty} = 2 - \log(\tau_\infty)$ (which represents the limiting Arrhenius slope at $T_g$, similar to the parameter used in refs.[28, 29]), it is easy to see that $B = m_P^{\infty}$, and eq.(3) is rewritten as $\log[\tau(T,V)] = 2 - m_P^{\infty} + m_P^{\infty} \left(\frac{T_g v_g^\gamma}{T v^\gamma}\right)^D$. Calculating $m_V$, we obtain $D = m_V/m_P^{\infty}$, so that

$$\log[\tau(T,V)] = 2 - m_P^{\infty} + m_P^{\infty} \left(\frac{T_g v_g^\gamma}{T v^\gamma}\right)^{\frac{m_V}{m_P^{\infty}}} \qquad (8)$$

In figure 1 are shown the relaxation time for several glass formers versus $Tv^\gamma$ normalized by its value at the glass transition. This plot is similar to the more usual fragility plot, but fig. 1 is not restricted to just the *T*-dependence, but considers all conditions of *T* and $v$. It can be seen that the rapidity of approach to the glass transition is rather different for each material; in fact, in this plot the steepness index at the glass transition is equal to $m_V$.[23]     [*figure 1 around here*]

If $T_g$ and $v_g$ are known, equation (8) has only three free parameters ($\gamma$, $m_V$, $m_P^{\infty}$). By using eq.(7) relating $m_V$ to $\gamma$ we can reduce this to two parameters ($\gamma$, $m_P^{\infty}$):

$$\log[\tau(T,V)] = 2 - m_P^{\infty} + m_P^{\infty} \left(\frac{T_g^{1/1+\gamma} v_g^{\gamma/1+\gamma}}{T^{1/1+\gamma} v^{\gamma/1+\gamma}}\right)^{\frac{m_P^{\max}}{m_P^{\infty}}} \qquad (9)$$

Since $m_P^{\infty} = 2 - \log(\tau_\infty)$, where $\tau_\infty$ is the high temperature limiting value of $\tau$, we expect its value to be similar for many materials $m_P^{\infty} \sim 12$ (see also data herein).[26,39] Interestingly, this roughly corresponds to the value of log($\tau_\infty$) beyond which the behavior becomes Arrhenius (log($\tau(T_A)$)~-10.2).[40]

Furthermore, since $m_P^{max}$ is a number, it follows from eq.(9) and the assumption that $m_P^\infty$ is constant for all materials that all the data should rescale on a single universal curve when plotted versus $(T\upsilon^\gamma)^{1/1+\gamma}$ normalized by its value at the glass transition (figure 2).    [*figure 2 around here*]

In figure2 we see that the data do indeed scale onto a single master curve, with the exception of 3 materials: PDE, KDE and 1,2-PB, which deviate for short $\tau$. However, eq.(9) accurately describes the data for even these materials. We illustrate this in figure 3 for PDE, with the fit to eq.(9), shown as a solid line, obtained using the known values of $T_g$ =293.7 K (at ambient P) and $\upsilon_g$ =0.72867 cm$^3$g$^{-1}$. The values of $m_P^\infty$, $m_P^{max}$ and $\gamma$ are given in the figure caption.    [*figure 3 around here*]

To determine if the deviation in figure 2 is due to differences in the values of $m_P^\infty$ among materials, we plot in figure 4 the function

$$\frac{\log[\tau(T,V)]-2}{m_P^\infty}+1=\left(\frac{T_g^{1/1+\gamma}\upsilon_g^{\gamma/1+\gamma}}{T^{1/1+\gamma}\upsilon^{\gamma/1+\gamma}}\right)^{\frac{m_P^{max}}{m_P^\infty}} \tag{10}$$

using the values of $m_P^\infty$=2-log($\tau_\infty$) reported in the caption. The agreement now is quite good, with only KDE showing any appreciable deviation. The origin of this deviation is unclear, however KDE in general is peculiar, for example behaving as a fragile liquid for some properties but as an intermediate liquid for others.[41] Its molecular structure is very close to that of PDE, but the temperature behavior is markedly different.[42]

In figure 4 we also show (solid line) the function $g(T,\upsilon)=\left(T_g^{1/1+\gamma}\upsilon_g^{\gamma/1+\gamma}/T^{1/1+\gamma}\upsilon^{\gamma/1+\gamma}\right)^{22}$ which describes well the master curve. From the mean value of $m_P^\infty$, we estimate $m_P^{max}=22\times m_P^\infty=22\times12\pm0.8=263\pm18$, which is consistent with the value of $m_P^{max}$ determined from the correlation between $m_V$ and $m_P$.[24] [*figure 4 around here*]

From the scaling observed in figure 4 it is evident that the parameters $m_P^\infty$ and $m_P^{max}$ are about constant for different materials. These quantities represent the respective activation energies at the glass transition normalized by the thermal energy in two limiting conditions. In fact, $m_P^\infty = 2-\log(\tau_\infty)$ is related to the activation energy when the thermal energy is much higher than the intermolecular barriers, so that any differences in molecular details becomes a second order effect. Note that the value for the "analogous" $m_P^\infty$ reported in the literature from fits using the Vogel-Fulcher-Tamman (VFT) equation[43-45] are generally larger. This is a consequence of the inability of the VFT equation to fit simultaneously data close to $T_g$ and at short relaxation times.[23,38,46-49] In fact, extrapolation of VFT fits near $T_g$ to short times predicts $\tau$ shorter (larger $m_P^\infty$) than the values actually measured.[38,48,49]

The parameter $m_P^{max}$ represents the maximum fragility, in the limit $\gamma \to 0$, in which case the relaxation time is a function of temperature alone. This corresponds to totally jammed dynamics, such that motion only occurs cooperatively, involving rearrangement of many neighboring molecules. The activation energy barrier is much larger than the thermal energy. In this limit the dynamics depends only on $T$ and $m_V = m_P$. The very high activation energy therefore reflects not a single energy barrier but rather the contribution from many different barriers. When the activation energy becomes so large relatively to the thermal energy, the molecular details become a second order effect so that an average value of $m_P^{max}$ gives a fair representation of the behavior. [In this case the motion is a cumulative result of many activated jumps, whereby the effective activation energy becomes very high even though the single jump barriers may be small. Small changes in the intermolecular barrier (i.e., those due to changes in volume) will be minimized. This interpretation reconciles the observation that, for a single material, the effect of temperature becomes more important close to $T_g$ as the motion becomes more cooperative.

The universal behavior in figure 4 clearly shows that the different dynamics (fragility) of glass-formers reflects primarily $\gamma$. There is a strong relationship between the volume-sensitivity of the relaxation times and their temperature sensitivity, because the volume-dependence reflects the nature of the intermolecular potential, which in turn determines the temperature dependence. Thus, fundamentally γ, while used to describe the volume dependence of τ, effectively describes the intermolecular potential, in particular the steepness of the repulsive interactions.

To summarize, eq.(1) has been firmly established by different experimental techniques. And a recent paper[23] showed the existence of a correlation between the isochoric, $m_V$, and the isobaric, $m_P$, fragilities, with the former moreover correlated with the parameter γ. Combining the idea that τ(T,V) approaching $T_g$ is related to γ, with the fact that the high temperature limiting dynamics of glass-formers is approximately universal, the inference is that the relaxation behavior *in toto* can be characterized by γ. Herein, these ideas are quantified by incorporating the correlations between $m_V$ and $m_P$ and $m_V$ and γ in eq.(8) to show that indeed the various behaviors are governed mainly by γ. A universal law (eq.(9)) is found, which is able to describe the evolution of the dynamics with changing temperature and density for all polymeric and molecular glass-formers in the range from very short times up to the glass transition. This law has 3 parameters (γ, $m_P^{max}$, $m_P^{\infty}$) which depend on the material, but only γ show significant variation, whereas $m_P^{max}$, $m_P^{\infty}$ are almost constant. The parameter γ takes into account the different effects of *T* and *V* on the dynamics, so that once the material-specific quantities related to γ have been normalized (see function $g(T,\upsilon)$), a universal power law is obtained, which describes the dynamics of glass-forming systems. A noteworthy consequence of this universal behavior is the absence of any divergence in the description of the glass transition; that is, no underlying transition need be invoked in order to interpret the slowing down of the dynamics.

**Figure captions**

**Figure 1.** The relaxation time for representative glass formers plotted versus the variable $T\upsilon^\gamma$ normalized by its value at the glass transition. The abbreviations in the caption are: 1,1'-di(4-methoxy-5-methylphenyl)cyclohexane (BMMPC),[30] 1,1'-bis(p-methoxyphenyl)cyclohexane (BMPC),[9] cresolphthalein-dimethylether (KDE),[41] (OTP/OPP),[31] phenyl salicylate (salol),[32] phenylphthalein-dimethylether (PDE),[38] poly(methyltolylsiloxane) (PMTS),[33] propylene carbonate (PC),[34] poly(phenyl glycidyl ether)-*co*-formaldehyde(PPGE),[35] 1,2 polybutadiene (1,2PB),[36] chlorinated biphenyl 42 (PCB42),[37] 54% chlorinated biphenyl (PCB54),[37] chlorinated biphenyl 62 (PCB62)[37].

**Figure 2.** The relaxation time for the same glass formers in figure 2, plotted versus the variable $\left(T\upsilon^\gamma\right)^{1/1+\gamma}$ normalized by its value at the glass transition.

**Figure 3.** Relaxation times for PDE[38] versus specific volume. Data correspond to an isobar at atmospheric pressure and five isotherms at the indicated temperatures. The solid lines are fits to eq.(9), taking $T_g$=293.7K and $\upsilon_g$=0.72867cm$^3$g$^{-1}$, which yields $m_P^\infty = 11.4 \pm 0.1$, $m_P^{max} = 264 \pm 1$ and $\gamma$ =4.37±0.01.

**Figure 4.** The relaxation time divided by the estimated value of $m_P^\infty$ for each material; $m_P^\infty$ = 12.35 for BMMPC, 12 for BMPC, 11.3 for KDE, 12.4 for OTP/OPP, 12.65 for salol, 11.5 for PDE, 12.7 for PMTS, 12.3 for PC, 12 for PPGE, 9.71 for 1,2PB, 12.2 for PCB42, 12.2 for PCB54, 12.2 for PCB62. The solid line corresponds to the function $g(T,\upsilon)$ defined in the text.

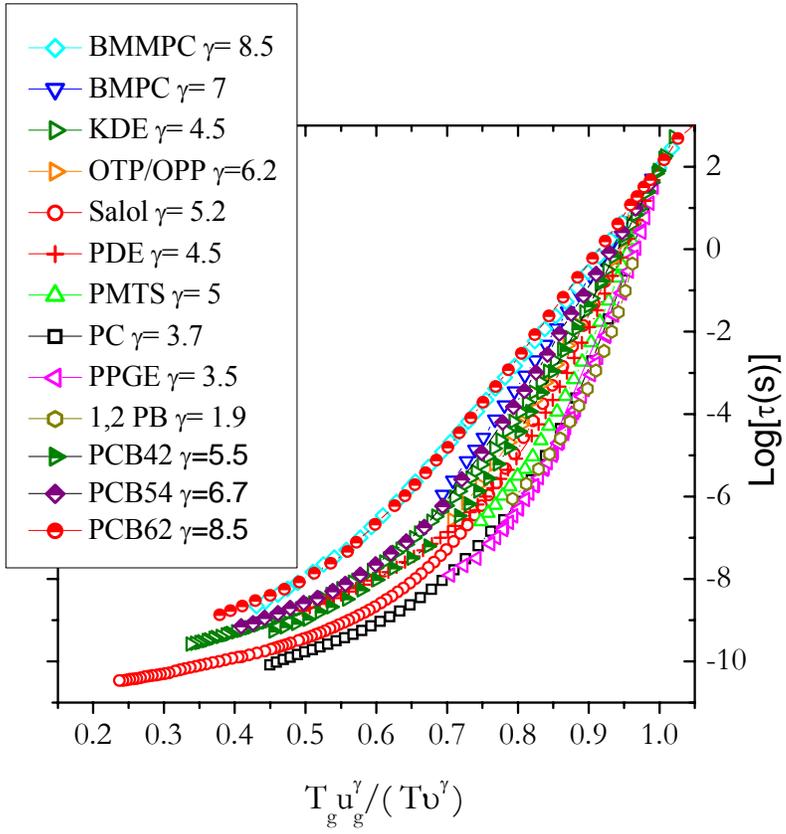

Figure 1

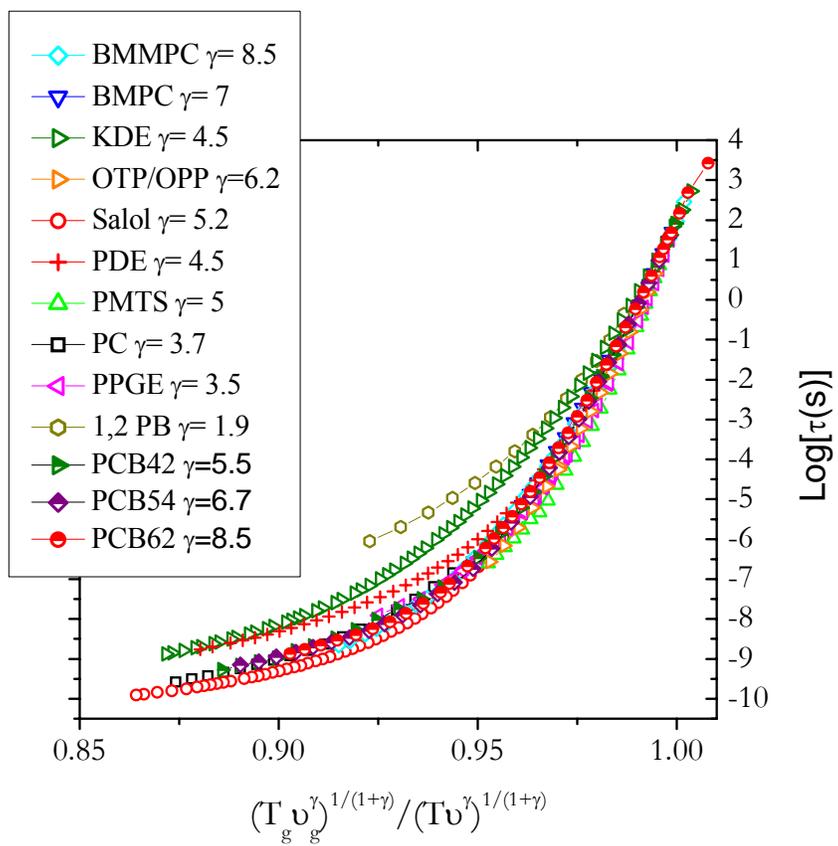

Figure 2

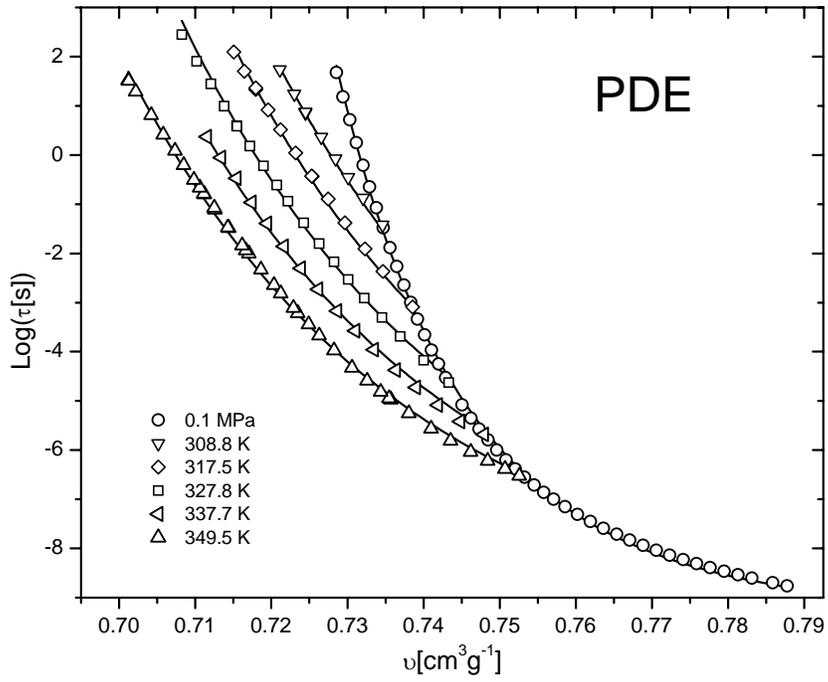

Figure 3

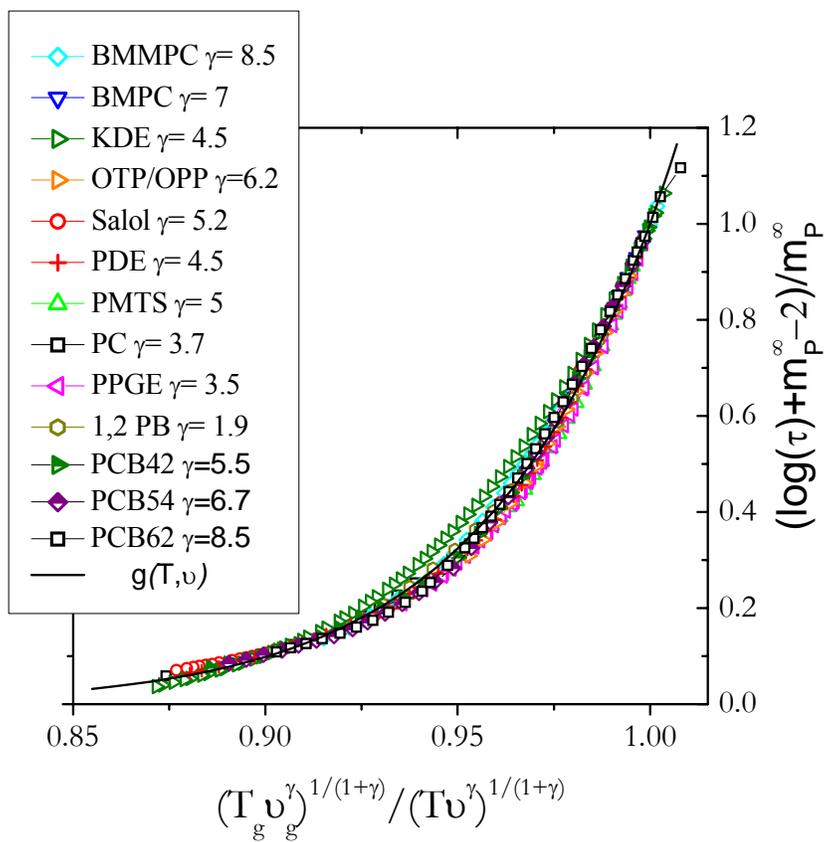

Figure 4